\documentclass[12pt]{article}

\pdfoutput=1

\usepackage[top=80pt,bottom=85pt,left=58pt,right=58pt]{geometry}
\usepackage{amssymb,amsmath,xypic}
\usepackage{slashed}
\usepackage{graphicx,color,subfigure}
\usepackage{float}
\usepackage{sectsty}
\usepackage{cite}

\usepackage[debug,pageanchor=false]{hyperref}
\definecolor{link}{rgb}{.8,.15,.1}
\hypersetup{colorlinks=true,linkcolor=link,citecolor=link,urlcolor=link,linktocpage}

 \allowdisplaybreaks

\makeatletter
\@addtoreset{equation}{section}
\makeatother

\newcommand{\beq}{\begin{equation}}
\newcommand{\eeq}{\end{equation}}
\newcommand{\bea}{\begin{eqnarray}}
\newcommand{\eea}{\end{eqnarray}}
\newcommand{\nn}{\nonumber}

\begin{document}
\begin{titlepage}

\begin{center}

\vskip .5in 
\noindent

{\Large \bf{ AdS$_2$ from Type IIA on weak G$_2$ manifolds }}

\bigskip\medskip

Niall T. Macpherson$^{a,b}$\footnote{macphersonniall@uniovi.es}, 
Achilleas Passias$^{c}$\footnote{achilleas.passias@uoi.gr} \\

\bigskip\bigskip\bigskip
{\small 

a:  Department of Physics, University of Oviedo, C/ Leopoldo Calvo Sotelo, 18, 33007 Oviedo, Asturia}

\medskip
{\small and}

\medskip
{\small 

b: Instituto Universitario de Ciencias y Tecnolog\'ias Espaciales de Asturias (ICTEA),\\
Calle de la Independencia 13, 33004 Oviedo, Spain}

\medskip
{\small 

c: Department of Physics, University of Ioannina, GR45110, Ioannina, Greece }
	
\bigskip\medskip

{\bf Abstract }
\end{center}
We present and study supersymmetric warped ${\rm AdS}_2 \times {\rm M}_7 \times I$ solutions of Type IIA supergravity, where M$_7$ is a weak G$_2$ holonomy manifold. Locally, the solutions are characterized by a cubic polynomial. We distinguish two branches, Branch 1 and Branch 2, and analyze their global properties, taking flux quantization into account. At the endpoints of the interval $I$ they exhibit either OF1-plane or O8/D8 type of singularities with Branch 1 allowing also for a Spin(7) cone. In the case of zero Romans mass, Branch 1 uplifts to eleven dimensions to ${\rm AdS}_4 \times {\rm M}_7$ whereas Branch 2 uplifts to warped ${\rm AdS}_3 \times {\rm M}_7 \times I$ with OM2-plane type of singularities.
		
\vskip .1in
	
\noindent 

\vskip .1in

\noindent

\noindent

\vfill
\eject

\end{titlepage}

\tableofcontents

\section{Introduction}
AdS$_2$ spacetimes are ubiquitous in string theory. A selection of recent developments includes brane intersections \cite{Lozano:2021rmk, Ramirez:2021tkd, Conti:2023naw}, defects \cite{Lozano:2021fkk, Lozano:2022vsv, Lozano:2022swp, Arav:2024wyg,Conti:2025wwf,Conti:2025wyj}, interfaces \cite{Chen:2021mtn, Gutperle:2024yiz}, branes on curved manifolds \cite{Ferrero:2020twa,Couzens:2021rlk,Couzens:2022yiv,Couzens:2022lvg, Faedo:2024upq}, scale-separation \cite{Tringas:2026ptt, Cribiori:2024jwq,Cribiori:2026qfs} and SL(2) T-duality \cite{Ramirez:2021tkd,Conti:2023naw,Conti:2023rul,Bielli:2024wru,Rojas:2026itv}. Supersymmetric AdS$_2$ solutions in 10 and 11 dimensions have been classified in terms of G-structures in \cite{Hong:2019wyi,Legramandi:2023fjr}. 

In \cite{Dibitetto:2018gbk} it was shown that the D0-F1-D8-brane configuration of \cite{Imamura:2001cr} admits an ${\rm AdS}_2 \times {\rm S}^7 \times I$ family of solutions as near-horizon geometries\footnote{It was also shown that this family of solutions can be obtained as the double analytic continuation of the ${\rm AdS}_7 \times {\rm S}^2 \times I$ family of solutions of \cite{Apruzzi:2013yva,Apruzzi:2015wna}.}. $I$ is a semi-infinite interval and the supersymmetry preserved is $\mathcal{N} = 8$. This family of AdS$_2$ solutions was revisited in \cite{Lozano:2025ief}, where $I$ was made finite by inclusion of D8-brane sources and the S$^7$ was replaced by its orbifold S$^7/\mathbb{Z}_k$, breaking supersymmetry to $\mathcal{N}=6$. A connection via T-duality to ${\rm AdS}_2 \times \mathbb{CP}^3 \times {\rm S}^1\times I$ in Type IIB was also established and proposals for the dual superconformal quantum mechanics were put forward. In \cite{Legramandi:2023fjr, Macpherson:2025aau}, where a larger study of Type II supersymmetric AdS$_2$ solutions was carried out, it was discovered that the ${\rm AdS}_2 \times {\rm S}^7 \times I$ family is part of a larger class of ${\rm AdS}_2 \times {\rm M}_7 \times I$ solutions where M$_7$ is a weak G$_2$ holonomy manifold and which is governed by  two functions subject to two ordinary differential equations. Generically, this class preserves $\mathcal{N}=1$ supersymmetry. Whenever one of the two aforementioned functions is constant, two branches of solutions appear, one of which includes the solutions of \cite{Dibitetto:2018gbk, Lozano:2025ief} with ${\rm M}_7 = {\rm S}^7/\mathbb{Z}_k$. We will refer to the latter as Branch 1 and to the second branch as Branch 2.

Both branches are characterized by a cubic polynomial which depends on the coordinate parameterizing $I$; the coefficient of the cubic term is proportional to the Romans mass. Both branches have electric NSNS 3-form and RR 2-form fields. Branch 2 has, in addition, a magnetic RR 4-form field. In contrast to Branch 1 which requires the addition of D8-branes for $I$ to be finite, the solutions of Branch 2 are compact before the addition of D8-brane sources. At the endpoints of the interval $I$, Branch 2  exhibits either a D8/O8 type of singularity or an OF1-plane type of singularity whereas Branch 1 allows also for a Spin(7) cone over M$_7$. The massless limit of Branch 1 uplifts to AdS$_4 \times$ M$_7$ in eleven-dimensional supergravity \cite{Dibitetto:2019nyz}, whereas for Branch 2 we find that it uplifts to warped AdS$_3$ solutions with OM2-plane \cite{Hanany:2000fq} singularities. 

The rest of the paper is organised as follows. In section \ref{branch1} we provide an analysis of Branch 1 of the ${\rm AdS}_2 \times {\rm M}_7 \times I$ solutions which includes their global properties, taking flux quantization into account. Section \ref{branch2} repeats the analysis for Branch 2 featuring examples with different physical characteristics. In Appendix \ref{roots} we look at the roots of the quartic polynomial that appears in the local expression of Branch 2 and Appendix \ref{uplift} includes the uplift to eleven dimensions of the massless limit of Branch 2.

\section{Branch 1}\label{branch1}
The first branch we will explore is a simple generalisation of the locally ${\rm AdS}_2\times{\rm S}^7\times I$ solutions first introduced in \cite{Dibitetto:2018gbk}, for which S$^7$ is replaced by a weak G$_2$ holonomy manifold M$_7$. M$_7$ comes with an associative three-form $\Phi_3$ with a single non-vanishing torsion class such that
\begin{equation}
d\Phi_3 = 4 \star \Phi_3,
\end{equation}
implying that $\Phi_3$ is only co-closed, rather than both closed and co-closed as it would be for a G$_2$ holonomy manifold.

Local solutions of this form were shown to preserve at least ${\cal N}=1$ supersymmetry in \cite{Legramandi:2023fjr}, with an enhancement to ${\cal N}=8$ when ${\rm M}_7 = {\rm S}^7$. Their NSNS sector (metric, dilaton $\Phi$, 2-form potential $B$) takes the form
\begin{align}
\frac{ds^2}{L^2}&=\frac{\sqrt{h'' h^3}}{8\Lambda} ds^2(\text{AdS}_2) + \frac{1}{8}\sqrt{\frac{h''}{h}}dr^2 +  \sqrt{\frac{h}{h''}}ds^2({\rm M}_7),~~~\Lambda \equiv  2 h h''- (h')^2,\nn\\[2mm]
e^{-\Phi}&=\frac{\sqrt{\Lambda}}{c_0 L^3}\left(\frac{h''}{h}\right)^{\frac{3}{4}},~~~B=-\frac{L^2}{8\sqrt{2}}\left(r-\ell+\frac{h h'}{\Lambda}\right) \text{vol}(\text{AdS}_2),\label{eq:sol1}
\end{align}
where the metric on AdS$_2$ has unit radius, $(\ell,c_0,L)$ are constants and $h=h(r)$ with a prime denoting differentiation. The non-trivial  RR  fluxes can be concisely expressed as
\beq
F_0=\frac{4}{c_0 L^4}h''',~~~\hat F_2= \frac{1}{2\sqrt{2}c_0 L^2}\left((r-\ell)h'''-h''\right) \text{vol}(\text{AdS}_2),
\eeq 
where $\hat F_2$ is a Page flux, related to the standard gauge invariant flux $F_2$ as $\hat F_2=F_2-B F_0$. One locally has a solution whenever $(F_0,\hat F_2)$ are closed and one can easily verify that
\beq
dF_0=\frac{4}{c_0 L^4}dr h'''',~~~d\hat F_2= \frac{1}{2\sqrt{2}c_0 L^2}(r-\ell)h''''dr\wedge \text{vol}(\text{AdS}_2),\label{eq:bianchis}
\eeq 
meaning that in regular parts of a solution we should have
\beq
h''''=0, 
\eeq
making $h$ locally a degree 3 polynomial. Of course $h'''\sim F_0$, and $F_0$ can shift across D8-brane sources, so this statement need only be true piecewise in a global solution, as we discuss in due course. 

\subsection{Metric reality and positivity}\label{sec:metposs}
A well defined solution requires that the warp factors in the metric and the dilaton are real and positive; this requires
\beq
\Lambda \equiv 2 h h''- (h')^2>0,\label{eq:realitiycond}
\eeq
which clearly implies $h h''>0$. In regular patches of a global solution we can express $h$ as a degree 3 polynomial. Let us suppose that the solution is regular in the interval $r\in(\ell,\ell+1)$, for $\ell$ a positive integer, such that we can take
\beq
h=h_{\ell}= \alpha_{\ell}+ \beta_{\ell}(r-\ell)+\frac{1}{2}\gamma_{\ell}(r-\ell)^2+\frac{1}{6}\sigma_{\ell}(r-\ell)^3.\label{eq:localh}
\eeq 
Following \cite{Lozano:2024idt}, which considers similar AdS$_3$ solutions, we will assume that $h h''>0$ is satisfied as\footnote{One could of course assume $(h<0, h''<0)$ but this doesn't lead to substantially different physics, and one or the other choice must be made in a global solution.}
\beq
h>0,~~~ h''>0.
\eeq
Then we should have
\beq
\alpha_{\ell}>0,~~~\gamma_{\ell}>0,
\eeq
so that $(h,h'')$ are positive at $r=\ell$. Under this assumption, it is possible to show that \eqref{eq:realitiycond} is satisfied for the range  $\ell\leq r\leq\ell+1$ in two rather simple cases:
\begin{enumerate}
\item  \eqref{eq:realitiycond} holds at $r=\ell$ and $\sigma_{\ell}\geq 0$
\item  \eqref{eq:realitiycond} holds at $r=\ell+1$ and $\sigma_{\ell}\leq 0$.
\end{enumerate}
Falling outside these two cases does not mean \eqref{eq:realitiycond} is necessarily violated, they just give simple ways to ensure it is satisfied.

\subsection{Brane charges}\label{eq:branecharges}
We would like to define some charges in terms of the above fluxes; we will not repeat the ``electric charge'' analysis of \cite{Lozano:2025ief} and instead consider standard Page charges on compact spaces\textemdash we again take $h$ as in \eqref{eq:localh} for $r\in(\ell,\ell+1)$.

We should have that $2\pi F_0$, which is the charge of D8-branes, is quantised\textemdash we thus tune 
\beq
c_0 L^4= 8\pi~~~\Rightarrow~~~N^{(8)}_{\ell}= 2\pi F_0\bigg\lvert_{r\in (\ell,\ell+1)}= \sigma_{\ell}, \label{eq:defN8ell}
\eeq
such that we simply need $\sigma_{\ell}\in \mathbb{Z}$ to ensure this. The EOM for $H=dB$ in this case is
\beq
d(e^{-2\Phi}\star H)= F_0 F_8.
\eeq 
As such, we can define a Page flux for fundamental strings as 
\beq
\hat H_7= e^{-2\Phi}\star H-F_0 C_7,~~~ F_8=dC_7
\eeq
which is such that $d\hat H_7=0$ away from sources. We can define $C_7$ in the interval $r\in(\ell,\ell+1)$ as\footnote{This expression also holds at the loci of D8-brane sources if they are placed at $r=\ell$.}
\begin{align}
C_7&=C_7^{\ell}=\frac{(2\pi)^7}{\text{Vol}(\text{M}_7)}c^{(7)}_{\ell} \text{vol}(\text{M}_7),\nn\\[2mm]
c^{(7)}_{\ell}&\equiv-\frac{3L^4 \text{Vol}(\text{M}_7)}{2c_0(2\pi)^7}\left(b_{\ell}+\frac{2h h'}{3h''}-2 (r-\ell)h+(r-\ell)^2h'-\frac{1}{3}(r-\ell)^3h''+\frac{1}{12}(r-\ell)^4 h'''\right),\label{eq:7formpotential}
\end{align}
where $b_{\ell}$ is an integration constant and the factors of $(\text{Vol}(\text{M}_7),(2\pi)^7)$ are chosen so that $\frac{1}{(2\pi)^7}\int_{\text{M}_7} C_7^{\ell}=c^{(7)}_{\ell}$. This leads to
\beq
\hat H_7=\frac{6}{c_0^2}\left(-\beta_{\ell}^2+2 \alpha_{\ell} \gamma_{\ell}+ b_{\ell} \sigma_{\ell}\right)\text{vol}(\text{M}_7).
\eeq
Upon introducing $Q$ as below and assuming the previous tuning for $c_0$, we find that the Page charge of fundamental strings is
\beq
N^{(1)}_{\ell}=  \frac{1}{(2\pi)^6}\int_{\text{M}_7|r\in(\ell,\ell+1)}\hat H_7= Q(-\beta_{\ell}^2+2\alpha_{\ell} \gamma_{\ell}+ b_{\ell} \sigma_{\ell}),~~~\frac{L^8}{2048 \pi^8}= \frac{Q}{3 \text{Vol}(\text{M}_7)}\label{eq:defN1ell}
\eeq
so we should have that $Q(-\beta_{\ell}^2+2\alpha_{\ell} \gamma_{\ell}+b_{\ell} \sigma_{\ell})\in\mathbb{Z}$ so that $N^{(1)}_{\ell}$ is likewise integer. The constant $Q$ is in principle redundant, but we will find it useful when we turn to the analysis of global solutions later. Note that we now have
\beq
c^{(7)}_{\ell}=-Q\left(b_{\ell}+\frac{2h h'}{3h''}-2 (r-\ell)h+(r-\ell)^2h'-\frac{1}{3}(r-\ell)^3h''+\frac{1}{12}(r-\ell)^4 h'''\right).
\eeq 
The final charge that is possible to define is for D0-branes, however this involves an integral over the entire internal space, and as such requires details of a global solution to perform. Nonetheless we can define
\beq
N^{(0)}=\frac{1}{(2\pi)^7}\int_{\text{M}_7\times I} F_8,
\eeq 
which should also be quantised when $r$ is bounded.

\subsection{Boundaries of $I$}\label{sec:bondaries1}
In order to construct a globally well defined solution with a compact internal space we need the interval $I$ to be bounded from above and below. So the question is: what are  the physical boundary behaviours that \eqref{eq:sol1} is consistent with? We will now address this. 

First off we should address the limit in which $F_0=0$ globally. In this case $h$ is only a degree 2 polynomial and it is possible to show that the resulting geometry is unique. In particular it is the IIA reduction of ${\rm AdS}_4/\mathbb{Z}_k\times{\rm M}_7$, for ${\rm M}_7$ any weak $G_2$ manifold, on the Hopf fiber U(1) of an ${\rm AdS}_3/\mathbb{Z}_k$ within ${\rm AdS}_4/\mathbb{Z}_k$\textemdash this was shown for the 7-sphere case in \cite{Lozano:2025ief}. To realise something more interesting we must allow $F_0\neq 0$, which we will now assume for the rest of this section.\footnote{Later we will consider global solutions with source D8-branes along the $r$ interval, for these $F_0=0$ is possible in local patches, just not globally.}

In order to bound the $r$ interval at one end we need the metric in \eqref{eq:sol1} to  tend to either a regular point or some singularity with a string theory origin, at some point $r=r_0$. Given the  topology of the metric, and which fluxes have compact support, the only plausible boundary behaviours in this case are the origin of a Spin(7) cone of base M$_7$, a D8/O8 system, F1-strings or OF1-planes. Following the analysis of \cite{Lozano:2025ief} it turns out that only 3 of these options are possible.  Given the invariance of local solutions under $r\to r+r_0$ we can always assume that a given boundary is at $r=0$. Under this assumption the possible boundaries, and how $h$ needs to be tuned to realise them in a local patch around $r=0$, are
\begin{align}
\text{Spin(7) cone}:& \quad h=(c_1+c_2 r)r^2,\nn\\[2mm]
\text{D8/O8}:& \quad h=c_1 +c_2 r^3,\nn\\[2mm]
\text{OF1}:& \quad h=\frac{c_1^2}{4c_2}+c_1 r+c_2 r^2+c_3 r^3, \label{eq:htunings}
\end{align}
where the constants $c_i$ must all be non-vanishing and be tuned such that \eqref{eq:realitiycond} is satisfied.

Given \eqref{eq:realitiycond}, it is possible to establish that each of the choices for $h$ in \eqref{eq:htunings} bound the interval as either  $r\in [0,\infty)$ or  $r\in (-\infty,0]$. As such, these profiles alone do not lead to bounded $I$; for that we need to glue local patches together with D8-branes along the interior of $r$ so that the space begins and ends at a physical boundary.

\subsection{D8-brane sources}\label{sec:sourceD8s}
A necessary ingredient for a bounded internal space are D8-branes along the interior of $r$. These allow us to glue local solutions together as, unlike the majority of string theory sources, D8-branes (or any effectively co-dimension 1 branes) don't bound the metric. They do however allow us to glue local patches with different values for $h$ together provided that the metric and dilaton, but not their derivatives, remain continuous. 

The Romans mass in a massive IIA solution need only be constant piecewise with discontinuities signalling the loci of D8-branes. Given that
\beq
F_0=\frac{4}{c_0 L^4}h''',
\eeq
this requires that $h'''$ is also  discontinuous across a D8-brane such that, for a D8-brane stack at $r=\ell$ we have
\beq
h''''=\Delta\sigma_{\ell}\delta(r-\ell),
\eeq
where $\Delta\sigma_{\ell}$ is the difference in D8-brane charge between $r=\ell+\epsilon$ and $r=\ell-\epsilon$. Tuning $(c_0,L)$ as before and taking $h$ to be tuned as \eqref{eq:localh} for $r\in(\ell,\ell+\epsilon]$ and \eqref{eq:localh} with $\ell\to \ell-1$ for  $r\in[\ell-\epsilon,\ell)$ this leads to 
\beq
dF_0\bigg\lvert_{r\sim \ell}=-\frac{1}{2\pi} (\sigma_{\ell-1}-\sigma_{\ell}) \delta(r-\ell)dr,
\eeq
which is consistent with a stack of $\sigma_{\ell-1}-\sigma_{\ell}$ D8-branes at $r=\ell$. Of course given that the D8-brane worldvolume is 9-dimensional it essentially couples to every other IIA field and can thus induce sources for all of them, at least in principle. Notice however that \eqref{eq:bianchis} implies that $d\hat F_2=0$ at $r=\ell$, the right-hand side vanishing as  $(r-\ell)\delta(r-\ell)\to 0$. As such, we conclude that the worldvolume flux of the D8-brane is trivial and the D8-branes do not form a bound state with D6-branes. Likewise, it is a simple matter to confirm that the Bianchi identities of the higher RR fluxes $(F_8=-\star F_2,F_{10}=\star F_0)$ do not receive source corrections.  The equation of motion of $H$ on the other hand does receive a source correction
\beq
d(e^{-2\Phi}\star H)- F_0 F_8\bigg\lvert_{r\sim \ell} \sim -(\sigma_{\ell-1}-\sigma_{\ell}) \delta(r-\ell)dr\wedge\text{vol}(\text{M}_7),
\eeq
but this comes about from the coupling of the D8-brane worldvolume to $B$\textemdash so is not signalling F1-string sources. 

Together the above implies that the local solutions are consistent with $\sigma_{\ell-1}-\sigma_{\ell}$ D8-branes at $r=\ell$, with zero worldvolume flux when $h'''$ experiences a shift at this locus. For this to be consistent the metric and dilaton should be continuous across the D8-brane which requires the same for
\beq
(h,~(h')^2, h'').
\eeq
Taking $h=h_{\ell}$ for $r\in(\ell,\ell+1)$ and $h=h_{\ell-1}$ for $r\in(\ell-1,\ell)$ with $h_\ell$ defined as in \eqref{eq:localh}, this implies that on either side of the D8-brane the constants $(\alpha_{\ell},\beta_{\ell},\gamma_{\ell})$ must be matched as
\begin{align}
\gamma_{\ell}&=\gamma_{\ell-1}+\sigma_{\ell-1},\nn\\[2mm]
\beta_{\ell}&=\pm\left(\beta_{\ell-1}+\gamma_{\ell-1}+\frac{1}{2}\sigma_{\ell-1}\right),\nn\\[2mm]
\alpha_{\ell}&=\alpha_{\ell-1}+\beta_{\ell-1}+\frac{1}{2}\gamma_{\ell-1}+\frac{1}{6}\sigma_{\ell-1}\label{eq:continuityconds},
\end{align}
with the $\pm$ following from $h'\to \pm h'$ as the D8-brane is traversed.  If these conditions are met such that \eqref{eq:realitiycond} holds and $\sigma_{\ell-1}-\sigma_{\ell}> 0$, then we have a well defined D8-brane source at $r=\ell$. However, \eqref{eq:continuityconds} has consequences for the Page charges on either side of the D8-brane.

First off, let us take $C_7$ as in \eqref{eq:7formpotential} for $r\in(\ell,\ell+1)$, with an analogous form with $\ell\to\ell-1$ for $r\in(\ell-1,\ell)$. Given this we find the following relation between the F1-string charges in adjacent cells
\beq
N^{(1)}_{\ell}=N^{(1)}_{\ell-1}+Q\left( b_{\ell}\sigma_{\ell}+\left(2\alpha_{\ell-1}+\beta_{\ell-1}+\frac{1}{3}\gamma_{\ell-1}+\frac{1}{12}\sigma_{\ell-1}-b_{\ell-1}\right)\sigma_{\ell-1}\right),
\eeq
which is insensitive to the sign in \eqref{eq:continuityconds}. It follows that $N^{(1)}_{\ell-1}$ integer implies $N^{(1)}_{\ell}$ integer provided that the final term in the above expression is also integer.

As we cross a D8-brane, $C_7$ must be continuous up to a large gauge transformation and this constraint depends on the sign taken in \eqref{eq:continuityconds}. Taking the $+$ sign in \eqref{eq:continuityconds} means that as we cross $r=\ell$
\beq
n^+_{\ell}=\bigg[\frac{1}{(2\pi)^7}\int_{\text{M}_7}C_7\bigg]_{r= \ell-\epsilon}^{r= \ell+\epsilon}=-Q (b_{\ell}-b_{\ell-1}+2\alpha_{\ell-1}+\beta_{\ell-1}+\frac{1}{3}\gamma_{\ell-1}+\frac{1}{12}\sigma_{\ell-1}),
\eeq
should be integer. If we instead take the $-$ sign in \eqref{eq:continuityconds} at $r=\ell$ we find that
\begin{align}
n^-_{\ell}&=\bigg[\frac{1}{(2\pi)^7}\int_{\text{M}_7}C_7\bigg]_{r= \ell-\epsilon}^{r= \ell+\epsilon}\\[2mm]
&= -Q\left(b_{\ell}-b_{\ell-1}+ \frac{1}{9}\left(12 \alpha_{\ell-1}+\beta_{\ell-1}-\gamma_{\ell-1}-\frac{1}{4}\sigma_{\ell-1}\right)-\frac{2(2\beta_{\ell-1}+\gamma_{\ell-1})(3(\alpha_{\ell-1}+\beta_{\ell-1})+\gamma_{\ell-1})}{9(\gamma_{\ell-1}+\sigma_{\ell-1})}\right)\nn.
\end{align}
This is clearly a more complicated expression to make integer than $n^+_{\ell}$, but quantisation of $n^-_{\ell}$ can still be achieved. Note that
\beq
n^-_{\ell}=n^+_{\ell}+\frac{Q\left(2\beta_{\ell-1}+2\gamma_{\ell-1}+\sigma_{\ell-1}\right)\left(6(\alpha_{\ell-1}+\beta_{\ell-1})+3\gamma_{\ell-1}+\sigma_{\ell-1}\right)}{9(\gamma_{\ell-1}+\sigma_{\ell-1})},\label{eq:nminusform}
\eeq
and irrespective of the sign in \eqref{eq:continuityconds}, one can show that
\beq
N^{(1)}_{\ell}=N^{(1)}_{\ell-1}-n^+_{\ell}\sigma_{\ell-1}-Q b_{\ell}(\sigma_{\ell-1}-\sigma_{\ell}),
\eeq
so that $N^{(1)}_{\ell}$ is guaranteed to be integer when $N^{(1)}_{\ell-1}$ is and $(n^+_{\ell},~Qb_{\ell}(\sigma_{\ell-1}-\sigma_{\ell}))$ are. 

\subsection{Global solutions}
As previously commented upon, to realise a global solution we need to include D8-branes along the $r$ interval in order to connect two of the possible physical boundary behaviours in  \eqref{eq:htunings}. There are numerous ways to achieve this in a well-defined way as long as certain constraints are obeyed; let us now study these. 

During the analysis of this section we shall assume that the $r$ interval is segmented into cells of length 1 such that in the $\ell$'th cell, where $r\in(\ell,\ell+1)$,  we have $h=h_{\ell}$. We will also assume the space begins in the $\ell=0$'th cell at $r=0$, \textit{i.e.} 
\begin{align}
h=\left\{ \begin{array}{l}\alpha_{0}+ \beta_{0}r+\frac{1}{2}\gamma_{0}r^2+\frac{1}{6}\sigma_{0}r^3~~~~0<r<1\\[2mm]
\alpha_{1}+ \beta_{1}(r-1)+\frac{1}{2}\gamma_{1}(r-1)^2+\frac{1}{6}\sigma_{1}(r-1)^3~~~~1<r<2\\[2mm]
\dots\\[2mm]
\alpha_{\ell}+ \beta_{\ell}(r-\ell)+\frac{1}{2}\gamma_{\ell}(r-\ell)^2+\frac{1}{6}\sigma_{\ell}(r-\ell)^3~~~~\ell<r<\ell+1\end{array}\right. .
\end{align}
We also take the earlier $\ell$-dependent expressions such as $(B,C_7)$ to define their values in the $\ell$'th cell with $r\in (\ell,\ell+1)$. At the points $r=\ell$ where the regular patches intersect we allow for possible D8-brane sources of charge
\beq
\sigma_{\ell-1}-\sigma_{\ell}\geq 0,
\eeq 
with equality allowed to cover the possibility that there is no such source.  If we assume that there are D8-branes located at $r=\ell$ for $\ell=1,\dots,\ell_{\text{max}}\in\mathbb{N}$, with no D8-branes for $r>\ell_{\text{max}}$, and that the space terminates at some $r=r_{\text{max}}>\ell_{\text{max}}$, it is then a fairly simple matter to confirm that the charge of D0-branes is given by
\beq
N^{(0)}=c^{(7)}_{\ell_{\text{max}}}(r=r_{\text{max}})-c^{(7)}_0(r=0)-\sum_{i=1}^{\ell_{\text{max}}}n_i,  
\eeq
where $n_i$ can be either $n_i^{\pm}$ depending on how \eqref{eq:continuityconds} is solved at each D8-brane. As these should be integer anyway, $N^{(0)}$ being integer boils down to the first 2 terms\textemdash note that $b_{\ell}$ drops out of $N^{(0)}$.

At $r=0$ we need to arrange for the space to begin with one of the boundary behaviours in \eqref{eq:htunings}. This amounts to demanding that the constants in the $0$'th cell are tuned as one of
\begin{align}
\text{Spin(7) cone}:& \quad \alpha_0=\beta_0=0,\nn\\[2mm]
\text{D8/O8}:& \quad \beta_0=\gamma_0=0,\nn\\[2mm]
\text{OF1}:& \quad 2\alpha_0\gamma_0=\beta_0^2 \label{eq:htunings2}.
\end{align}
In each case the metric is well-defined up to the first D8-brane provided that whichever of $(\alpha_0,\gamma_0)$ are non-zero, are positive. One also has that
\beq
N^{(1)}_0= Q b_0 \sigma_0,
\eeq
irrespective of how one chooses to begin the space, so  $Q b_0\sigma_0\in\mathbb{Z}$ is required.

Next we need to ensure that 
\beq
\Lambda > 0\label{eq:psscond}
\eeq
holds across all cells except at the bounds of the $r$ interval, where it must vanish. As explained in section \ref{sec:metposs}, if this inequality holds at $r=\ell$ it is guaranteed to hold across the entire $\ell$'th cell if $\sigma_{\ell}\geq 0$. It then follows from the metric continuity constraints in \eqref{eq:continuityconds}, with $h'_{\ell-1}(\ell)=+ h'_{\ell}(\ell)$, that if \eqref{eq:psscond} held in the $(\ell-1)$'th cell it also holds in the $\ell$'th cell provided that $\sigma_{\ell}\geq 0$. Thus when \eqref{eq:psscond} holds in the $0$'th cell, it also holds across an arbitrary number of cells moving along $\mathbb{R}^+$ towards infinity provided that $\sigma_{\ell}\geq 0$ in each of them. 
If $h'_{\ell-1}(\ell)=+ h'_{\ell}(\ell)$ and $\sigma_{\ell}\geq 0$ are imposed across all cells up to and including the $(P-1)$'th cell ending at $r=P$, then we can express the parameters in some $\ell$'th of these cells, for $\ell=1,\dots,P-1$, in terms of those of the $0$'th cell as
\begin{align}
Q b_{\ell}&=Q\left(b_0-2 \ell \alpha_0-\ell^2\beta_0-\frac{\ell^3}{3}\gamma_{0}-\frac{1}{12}\sum_{i=1}^{\ell}(2i-1)(2i(i-1)+1)\sigma_{\ell-i}\right)-\sum_{i=1}^{\ell}n^+_{\ell+1-i},\nn\\[2mm]
\alpha_{\ell}&=\alpha_0+\ell \beta_0+ \frac{\ell^2}{2}\gamma_0+\frac{1}{6}\sum_{i=1}^{\ell}(3(i-1)i+1)\sigma_{\ell-i},\nn\\[2mm]
\beta_{\ell}&=\beta_0+\ell \gamma_0+\frac{1}{2}\sum_{i=1}^{\ell}(2i-1)\sigma_{\ell-i},\nn\\[2mm]
\gamma_{\ell}&=\gamma_0+\sum_{i=1}^{\ell}\sigma_{\ell-i}
. \label{eq:ellthcellintermsof0th}
\end{align}
Given the form of $N^{(1)}_{\ell}$ in \eqref{eq:defN1ell}, and where $\frac{1}{12}$, $\frac{1}{6}$, $\frac{1}{3}$ and $\frac{1}{2}$ lie in \eqref{eq:ellthcellintermsof0th} it is fairly obvious that if one tunes
\beq
Q b_{\ell}= m_{\ell},~~~(Q, \alpha_0,\beta_0,m_{\ell}) \in\mathbb{Z},~~~~ (2\gamma_{0},\sigma_{\ell})\in12\mathbb{Z},\label{eq:quantisation1}
\eeq
then everything that ought to be an integer between the $0$'th and $(P-1)$'th cell is indeed an integer.\footnote{We do not claim that this is the only way to achieve quantisation.}

We now need to make the space terminate at some upper bound of the $r$ interval $r=r_{\text{max}}$. The simplest way to achieve such a bounded solution is to impose that $\sigma_{\ell}>0$ and $h'$ does not change sign until one reaches some $P$'th cell. In this cell the sign of $h'$ does not change but one now imposes $\sigma_{P}<0$. This typically leads to an OF1-plane at $r=r_{\text{max}}>P$; anything else would require a fine tuning of the parameters which looks impossible for integer $\sigma_{\ell}$. However, $r_{\text{max}}$ is the solution to a degree 4 polynomial with coefficients that are non-linear combinations of the parameters  $(\alpha_0,\beta_0,\gamma_0,\sigma_\ell)$, so it seems very unlikely that one can tune $r_{\text{max}}=P+1$, or some other integer value. This is related to a problem for the flux quantisation of $N^{(0)}$, as the $c^{(7)}_P(r=r_{\text{max}})$ term will typically only be integer at very particular values of $r_{\text{max}}$, values for which the other parameters that should be integer are not\textemdash thus we need to do something a little less simple.

\begin{figure}[h]
\centering
\includegraphics[scale=0.6]{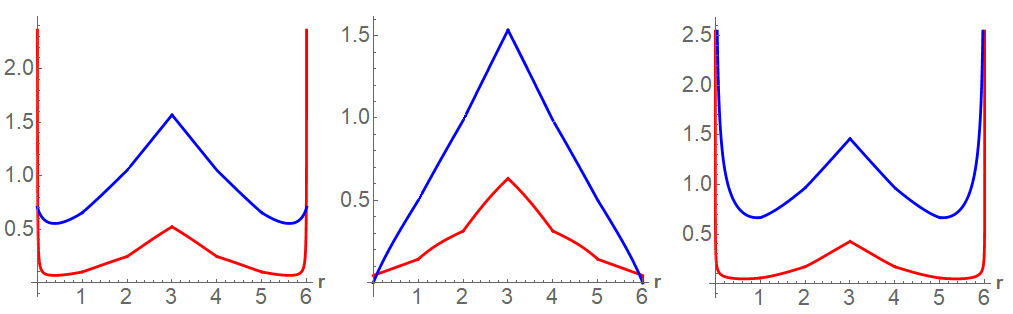}
\caption{In each case the blue curve is the radius squared of $\text{M}_7$ while the red is that of AdS$_2$, with respect to \eqref{eq:sol1} with $L=1$. In each case $P=3$  and $(\sigma_0,\sigma_1,\sigma_2)=-(\sigma_5,\sigma_4,\sigma_3)=(36,24,12)$. From left to right the tuning of the remaining parameters is $(\alpha_0,\beta_0,\gamma_0)=(3,6,6)$ bounding the space between two OF1-planes, $(\alpha_0,\beta_0,\gamma_0)=(0,0,12)$ bounding the space between the origin of two Spin(7) cones, $(\alpha_0,\beta_0,\gamma_0)=(10,0,0)$ bounding the space between two D8/O8 systems.}\label{fig:plots1}
\end{figure}

There is a particular tuning of the D8-brane charges and choice of signs in the continuity constraints that allows for symmetric solutions that begin and end with the same type of physical boundary behaviour. To achieve this one again keeps $\sigma_{\ell}\geq 0$ and the sign of $h'$ fixed across all D8-branes until one located at  $r=P$. At this D8-brane one allows $h'$ to change sign and identifies
\beq
\sigma_{P}=-\sigma_{P-1}.
\eeq
As $\sigma_{P}$ is now negative we must check if \eqref{eq:psscond} is satisfied at $r=P+1$; if it is, then we know it is satisfied across the entire $P$'th cell. Given the above identifications, $h$ in the $P$'th cell can be expressed in terms of the charges in the $(P-1)$'th cell as
\beq
h=  \alpha_{P-1}-\beta_{P-1}(r-(P+1))+\frac{1}{2}\gamma_{P-1}(r-(P+1))^2-\frac{1}{6}\sigma_{P-1}(r-(P+1))^3,
\eeq
as such at $r=P+1$ \eqref{eq:psscond} becomes
\beq
-\beta_{P-1}^2+2\alpha_{P-1}\gamma_{P-1}>0
\eeq
which matches the value at $r=P-1$ that we have already ensured is positive. If we now go back to solving \eqref{eq:continuityconds} without $h'$ changing signs at the D8-branes for $r=P+1$ and beyond, one finds that
\beq
\sigma_{P+n}=-\sigma_{P-n-1} ~~~ \Rightarrow ~~~ \alpha_{P+n}=\alpha_{P-n},~\beta_{P+n}=-\beta_{P-n},~\gamma_{P+n}=\gamma_{P-n},
\eeq
for $n=1,\dots,P-1$, and that \eqref{eq:psscond} is still satisfied until one reaches $r=2P$, where the space now terminates with the same behaviour as it started with at $r=0$.  The quantisation of $( n^-_P, N^{(1)}_P, N^{(8)}_{P})$  is guaranteed if one tunes the various constants as in \eqref{eq:quantisation1}, but further refines
\beq
Q= 3(\gamma_{P-1}+\sigma_{P-1}) q,~~~ q\in\mathbb{Z}, \label{eq:quantisation}
\eeq
which ensures that $n^-_P$ is an integer given\footnote{Given \eqref{eq:quantisation1} the numerator of the second term in \eqref{eq:nminusform} is a multiple of 12.}  \eqref{eq:nminusform}. For the remaining cells with $P+1 \leq r \leq2P$ it is possible to show that
\begin{align}
n^+_{P+n}&= n^+_{P-n+1}- m_{P+n}+m_{P+n-1}+(m_{P-n+1}-m_{P-n}),\nn\\[2mm]
N^{(1)}_{P+n}&=N^{(1)}_{P-n}-(m_{P+n}\sigma_{P-n-1}+m_{P-n}\sigma_{P-n})
\end{align}
for $n=1,\dots,P-1$, thus quantisation in these cells is ensured by our current assumptions. All that remains is to confirm that the charge of D0-branes is also quantised. To this end, one can establish that
\beq
c^{(7)}_{2P-1}(r=2P)-c^{(7)}_{0}(r=0)= -m_{2P-1}+2m_0-m_1- n_1^{+}+ u,
\eeq
where  $u$ is a constant that depends on the type of boundary behaviour we choose, \textit{i.e.}
\beq
u= \left\{ \begin{array}{l}  0,~~~ \text{D8/O8 or  Spin(7) cone}\\[5mm]
\displaystyle \frac{4Q \alpha_0\beta_0}{3 \gamma_0},~~~ \text{OF1}\end{array}\right.
\eeq
Thus we find that the D0-brane charge is 
\beq
N^{(0)}=  -m_{2P-1}+2m_0-m_1- n_1^{+}+ u-\sum_{i=1}^{P-1}n^+_i-n^-_P -\sum_{i=P+1}^{2P-1}n^+_i,
\eeq
where one should note that this is actually independent of all $m_{\ell}$. This is automatically integer under our current assumptions for solutions bounded between D8/O8 singularities or Spin(7) cones. When a solution is bounded between OF1-planes however we must additionally impose
\beq
u= \frac{4Q \alpha_0\beta_0}{3 \gamma_0}\in \mathbb{Z},
\eeq
to make $N^{(0)}$ integer. Plots of such symmetric solutions, that satisfy flux quantisation for $Q$ tuned as in  \eqref{eq:quantisation}, can be found in Figure \ref{fig:plots1}.

\section{Branch 2}\label{branch2}
We now consider a deformation of the local solutions in the previous section in which the weak G$_2$ holonomy 3-form supported by M$_7$ is allowed to appear in the fluxes. Such solution generically preserve ${\cal N}=1$ supersymmetry, even in the case that ${\rm M}_7={\rm S}^7$ as in this case $\Phi_3$ is charged under the SO$(8)$ of the 7-sphere. 

Following \cite{Legramandi:2023fjr}, the NSNS sector of the local solution is given by:
\begin{align}
\frac{ds^2}{L^2} &= \frac{\sqrt{12h^3h''}}{8\sqrt{5}\Delta}ds^2(\text{AdS}_2) + \frac{1}{8}\sqrt{\frac{5h''}{12h}}dr^2 + \frac{25}{36}\sqrt{\frac{12h}{5h''}} ds^2({\rm M}_7),~~~ \Delta \equiv 2 hh''-\frac{12}{5}(h')^2, \nn\\[2mm]
e^{-\Phi} &= \frac{18}{25}\left(\frac{3}{5}\right)^{\frac{1}{4}}\frac{\sqrt{\Delta}}{c_0 L^3}\left(\frac{h''}{h}\right)^{\frac{3}{4}},~~~B = -\frac{L^2}{8 \sqrt{2}}\left(r-\ell+\frac{12hh'}{5\Delta}\right) \text{vol}(\text{AdS}_2)
\label{eq:caseIImet}
\end{align}
where $(\ell,c_0,L)$ are constants and the metric on AdS$_2$ has radius 1. As with the solutions of the previous section $h$ is piecewise a polynomial of $r$ of degree 3. The RR fluxes can be expressed as
\begin{align}
F_0&=\sqrt{\frac{6}{5}}\frac{72}{25c_0 L^4} h''',\nn\\[2mm]
\hat F_2&=F_2-F_0 B=\sqrt{\frac{3}{5}}\frac{9}{25c_0 L^2}\left((r-\ell)h'''-h''\right)\text{vol}(\text{AdS}_2),\nn\\[2mm]
 F_4&=dC_3,~~~C_3=-\frac{1}{\sqrt{30}c_0 }h' \Phi_3.\label{eq:caseIIfluxes}
\end{align}
We now have a non-zero $F_4$ which has compact support when the internal manifold is compact.

\subsection{Metric reality and positivity and the range of the coordinate $r$}
Similarly to Branch 1, positivity of the metric requires the polynomial
\begin{equation}
\Delta \equiv 2 hh''-\frac{12}{5}(h')^2 \geq 0.
\end{equation}
However, ensuring this, is not simply a matter of requiring that the coefficients in a local expression for $h$ are positive.

First let us assume that $F_0 \neq 0$, which implies that $\Delta$ is a quartic polynomial. From the definition of $\Delta$ in terms of the cubic polynomial $h$, it follows that the coefficient of its $r^4$ term is negative (see Appendix \ref{roots}). This implies that for a physical solution the $r$ interval is always constrained to some finite portion of $\mathbb{R}$. In order for $\Delta$ to stay positive we need to take $r$ between two of its roots. An analysis of the roots of the quartic polynomial $\Delta$ (Appendix \ref{roots}) shows that in order for $\Delta$ to have two distinct real roots, its discriminant, Disc$(\Delta)$, has to be negative. It is the case that Disc$(\Delta) \propto$ Disc$(h)$ and so Disc$(h)$ cannot be zero, that is, $h$ cannot have multiple roots. This is in contrast with the Branch 1.

For the case of $F_0 = 0$, $\Delta$ is quadratic with negative $r^2$ coefficient. Hence again, metric positivity implies a finite $r$ interval. For $\Delta$  to be positive it must have two distinct real roots between which $r$ lies. This is achieved if the discriminant of $\Delta$ is positive.

\subsection{Brane charges}
We now turn our attention to flux quantisation: The fluxes with compact support are now $(F_0, F_4, F_8, \star H)$ which give rise to charges for D8-, D4- and D0- branes and F1-strings, however only the D8 and F1 charges do not require integrating over $r$. As  $F_2 \wedge F_6=0$, the equation of motion for the NSNS 3-form, away from possible sources, is
\beq
d(e^{-2\Phi}\star H)= \frac{1}{2}F_4\wedge F_4+F_0 F_8.
\eeq
As such 
\beq
\hat H_7= e^{-2\Phi}\star H-\frac{1}{2}C_3\wedge F_4- F_0 C_7,
\eeq
is closed and so suitable for defining a conserved charge. Again assuming that we are in a cell for which $r\in(\ell,\ell+1)$ we find that
\begin{align}
C_7&=C_{7}^{\ell} = \frac{(2\pi)^7}{\text{Vol}(\text{M}_7)} c^{(7)}_{\ell} \text{vol}(\text{M}_7), \label{eq:C72}\\[2mm]
c^{(7)}_{\ell}& \equiv -\frac{275\sqrt{30} L^4 \text{Vol}(\text{M}_7) }{ 1296 c_0 (2\pi)^7}\left(b_{\ell}+\frac{12 h h'}{11 h''}-2(r-\ell) h+(r-\ell)^2 h'-\frac{1}{3}(r-\ell)^3 h''+\frac{1}{12}(r-\ell)^4 h'''\right),\nn
\end{align}
is such that $dC_7=F_8$, where $b_{\ell}$ is a constant, and $\frac{1}{(2\pi)^7}\int_{\text{M}_7}C_7^{\ell}= c^{(7)}_{\ell}$. Then taking $C_3$ as defined in \eqref{eq:caseIIfluxes} and $h$  as in \eqref{eq:localh} we find that
\beq
\hat H_7= \frac{11}{3 c_0^2}(-\beta_{\ell}^2+2\alpha_{\ell}\gamma_{\ell}+b_{\ell}\sigma_{\ell})\text{vol}(\text{M}_7).
\eeq
Thus, we precisely reproduce the D8 and F1 charges $(N^{(8)}_{\ell},N^{(1)}_{\ell})$ in terms of constants $(\alpha_{\ell},\beta_{\ell},\gamma_{\ell},\sigma_{\ell},Q)$ as found earlier in \eqref{eq:defN8ell} and \eqref{eq:defN1ell} if we tune
\beq
c_0=\frac{144\pi}{25 L^4}\sqrt{\frac{6}{5}},~~~~L^8 =\frac{1}{11}\frac{3^6 2^{15}\pi^8 Q}{5^5 \text{Vol}(\text{M}_7)}.\label{eq:tuningLc02}
\eeq
It then follows that
\beq
c^{(7)}_{\ell}=-Q \left(b_{\ell}+\frac{12 h h'}{11 h''}-2(r-\ell) h+(r-\ell)^2 h'-\frac{1}{3}(r-\ell)^3 h''+\frac{1}{12}(r-\ell)^4 h'''\right).
\eeq
The D0 and D4 charges both require us to have a concrete global completion to compute, but we can still give their definitions, namely
\beq
N^{(0)}=\frac{1}{(2\pi)^7}\int_{\text{M}_7\times I}F_8,~~~~~N^{(4)}=-\frac{1}{(2\pi)^3}\int_{\Sigma_3\times I}F_4,\label{eq:N0N4defs}
\eeq
where we should stress that $N^{(4)}$ can only be defined if there exists a 3-cycle in M$_7$ that $F_4$ has support on\textemdash for instance S$^7$ has no such 3-cycle, while S$^7/\mathbb{Z}_k$ does.

\subsection{Boundaries of $I$}
In this section we examine the possible boundary behaviours for Branch 2 which define the possible global completions. As this analysis is new, we will be more explicit about what the fields of the local solution approach at these boundaries than in section \ref{sec:bondaries1}.

It is possible to establish that the only boundary behaviours that the local solution of Branch 2 is compatible with are a D8/O8 singularity and an OF1-plane extended in AdS$_2$ and localised at the tip of a Spin(7) cone. Both of these happen at a loci where $\Delta$ vanishes in the form of a simple root. Let us now see how these behaviours come about.

Near a simple root of $\Delta$, $r_0$, for which $h''\neq 0$ the metric at leading order becomes
\begin{align}
\frac{ds^2}{L^2}&\sim \frac{\sqrt{12h^3(r_0)h''(r_0)}}{8\sqrt{5}\Delta'(r_0)}(r-r_0)^{-1}ds^2(\text{AdS}_2)+\frac{25}{36}\sqrt{\frac{12h(r_0)}{5h''(r_0)}} ds^2({\rm M}_7)+ \frac{1}{8}\sqrt{\frac{5h''(r_0)}{12h(r_0)}}dr^2,
\end{align}
while the dilaton becomes
\begin{align}
e^{-\Phi}&\sim \frac{18}{25}\left(\frac{3}{5}\right)^{\frac{1}{4}}\frac{\sqrt{\Delta'(r_0)}}{c_0 L^3}\left(\frac{h''(r_0)}{h(r_0)}\right)^{\frac{3}{4}}(r-r_0)^{\frac{1}{2}}.
\end{align}
This is the singularity associated to a localised OF1-plane extended in AdS$_2$.

On the other hand, at a simple root of $\Delta$, $r_0$, for which $h', h''\to 0$ the metric at leading order becomes
\begin{align}
\frac{ds^2}{L^2}&\sim (r-r_0)^{-\frac{1}{2}} \sqrt{\frac{12h(r_0)}{5h'''}} \left( \frac{1}{16}ds^2(\text{AdS}_2)+\frac{25}{36}ds^2({\rm M}_7) \right) + (r-r_0)^{\frac{1}{2}} \frac{1}{8}\sqrt{\frac{5h'''}{12h(r_0)}}dr^2,
\end{align}
and the dilaton
\begin{align}
e^{-\Phi}&\sim \frac{18}{25}\left(\frac{12h'''}{5h(r_0)}\right)^{\frac{1}{4}}\frac{h'''}{c_0 L^3}(r-r_0)^{\frac{5}{4}}.
\end{align}
We recognise this singularity as an D8/O8 system extended in ${\rm AdS}_2 \times {\rm M}_7$.

We can of course use the invariance of the local solution under $r\to r+r_0$ to assume that a given boundary is at $r=0$. Under this assumption the two possible boundary behaviours require tuning
\beq
\text{D8/O8}:~  h= c_1+ c_2 r^3,~~~\text{OF1}:~ h= \frac{3c_1^2}{5c_2}+c_1 r+c_2 r^2+c_3 r^3, \label{eq: tuning2}
\eeq
where $c_i$ are constants that must be tuned such that the metric is positive. There are three notable differences between the possible boundaries in Branch 2 and Branch 1: i) The origin of the Spin(7) cone is no longer possible: one cannot make the M$_7$ warp factor vanish at a point where $\Delta$ is positive. ii) The OF1-plane now permits $c_3=0$ meaning that there no longer need be a Romans mass to realise this boundary. iii) Tuning  $h$ globally as in \eqref{eq: tuning2} does not lead to the $r$ interval being semi-infinite, as we will explore in the next section. 

\subsection{Global solutions with constant $F_0$}
In  contrast to Branch 1, in this Branch it is possible to bound the interval parameterised by $r$ while keeping $F_0$ constant globally, meaning that there are no D8-brane sources in the interior of the $r$ interval.

One can distinguish between three physically distinct options
\begin{itemize}
\item {\bf Case 1}. In this case the Romans mass is non-zero and $h$ is a cubic polynomial. The coefficients of $h$ are restricted so that the discriminant of $\Delta$ is negative. The range of $r$ is $[r_-,r_+]$ where $r_-$ and $r_+$ are roots of $\Delta$. At the endpoints of the $r$-interval there are OF1 type singularities.
\item {\bf Case 2}. In this case 
\begin{equation}
h = c_1 + c_2 (r-r_0)^3, 
\end{equation}
where $c_1$, $c_2$ are constants and the Romans mass is proportional to $c_2$. The range of $r$ is $[r_0,r_+]$ where $r_0$ and $r_+$ are roots of $\Delta$, with $r_0$ being also a double root of $h'$. Near $r_0$ there is an O8/D8 type singularity and near $r_+$ an OF1 type one.
\item {\bf Case 3}. In this case the Romans mass is zero and $h$ is a quadratic polynomial. The coefficients of $h$ are restricted so that the discriminant of $\Delta$ is positive. The range of $r$ is $[r_-,r_+]$ where $r_-$ and $r_+$ are the roots of $\Delta$. At the endpoints of the $r$-interval there are OF1 type singularities.
\end{itemize}
The third of these can also be lifted to $d=11$ supergravity which results in a solution bounded between to OM2-planes, as we explore in Appendix \ref{eq:appendix2}.

Let us now analyse flux quantisations for these three solutions. We will take
\beq
h= \alpha_0+ \beta_0 r+ \frac{1}{2}\gamma_0 r^2+\frac{1}{6} \sigma_0 r^3,
\eeq
and assume that $C_7$ is as in \eqref{eq:C72} for $\ell=0$. As such,  upon tuning $(L,c_0)$ as in \eqref{eq:tuningLc02}, the D8 and F1 charges are
\beq
N^{(8)}=\sigma_0,~~~ N^{(1)}=Q(-\beta_0^2+ 2 \alpha_0 \gamma_0+b_0 \sigma_0)\label{eq:thing}.
\eeq

We will without loss of generality assume that $r_0=r_-=0$. 

\subsubsection{Solution bounded between D8/O8 and OF1}
We can realise a solution between a D8/O8 system at $r=0$ and an OF1-plane at $r=r_+>0$ by tuning
\beq
\alpha_0= \frac{2}{15} r_+^3 \sigma_0,~~~\beta_0=\gamma_0=0,~~~\sigma_0\neq 0.
\eeq
This tuning is well defined irrespective of the sign of $\sigma_0$, so as long as this is integer, $N^{(8)}$ is likewise. We can compute the charge of D0-branes and F1-strings from \eqref{eq:N0N4defs} and \eqref{eq:thing}, which leads to
\beq
N^{(0)}=\frac{41}{220}Q r_+^4 \sigma_0,~~~N^{(1)}= Qb_0 \sigma_0
\eeq
where we clearly have the freedom to tune $(Qr_+^4, Qb_0)$ such that these are integer \textit{i.e.} we can use the above expression to eliminate $(Q,b_0)$ in favour of $(N^{(0)},N^{(1)})$. All that remains is the question of the quantisation of the D4-brane charge if M$_7$ has a 3-cycle which $\Phi_3$ has support on\textemdash such a cycle does not exist for all M$_7$,  S$^7$ lacks one for example. If we assume such a cycle $\Sigma_3$ does exist then \eqref{eq:N0N4defs} now implies that
\beq
N^{(4)}=\sqrt{\frac{N^{(0)}N^{(8)}}{82\text{Vol}(\text{M}_7)}}\int_{\Sigma} \Phi_3.
\eeq
A manifold with such a 3-cycle is S$^7/\mathbb{Z}_k$, with the orbifolding performed on the circle bundle in ${\rm S}^1\hookrightarrow{\rm S}^7\to\mathbb{CP}^3$\textemdash actually it has two of them. In this case one has
\beq
\text{Vol}(\text{S}^7/\mathbb{Z}_k)= \frac{\pi^4}{3k}.
\eeq
As exemplified by its parameterisation as a foliation by $T^{(1,1)}$ over an interval, $\mathbb{CP}^3$ contains two distinct copies of $\mathbb{CP}^1$; we label these $\mathbb{CP}^1_{\pm}$. The 3-cycles in question are then
\beq
\Sigma^{\pm}_3=\mathbb{CP}^1_{\pm}\times \text{S}^1.
\eeq
S$^7/\mathbb{Z}_k$ supports a $\Phi_3$ such that\footnote{Actually there are 6 independent $\Phi_3$ as in isolation S$^7/\mathbb{Z}_k$ preserves 6 supercharges, but only one can appear in the flux, and which is elected to appear makes no difference to what follows.}
\beq
\int_{\Sigma_3^{\pm}} \Phi_3 = \pm \frac{2\pi^2}{k}.
\eeq
In this case in particular we thus arrive at
\beq
N^{(4)}_{\pm}= \pm \sqrt{\frac{6 N^{(0)}N^{(8)}}{41 k}},
\eeq
which are clearly not generically integer, but can be made so by tuning $(N^{(0)},N^{(8)},k)$ appropriately. For instance, ($N^{(0)}= n_1 p^2$, $N^{(8)}=  n_2 q^2$, $n_1 n_2= 6\times 41 k$) for $n_{1,2}>0$, $p,q\in\mathbb{Z}$ achieves the desired goal, but there are additional choices. 

Thus flux quantisation can be achieved quite easily when M$_7$ contains no 3-cycle to integrate $\Phi_3$ on, but a more delicate tuning of the other charges is required when such a cycle does exist. This pattern will persist with the remaining examples.

\subsubsection{Solution bounded between OF1 and OF1 with $F_0=0$}
We have a solution in massless IIA with OF1-planes located at $r=0,r_+$ when we tune
\beq
\alpha_0= \frac{3 r_+^2}{10}\gamma_0,~~~\beta_0=-\frac{r_+}{2}\gamma_0,~~~\sigma_0=0,~~~\gamma_0\neq 0
\eeq
where the sign of $\gamma_0$ plays no role. Clearly we have $N^{(8)}=0$ and we can deal with the quantisation of the charge of F1-strings and D0-branes by  fixing 
\beq
Q= \frac{20 N^{(1)}}{7  r_+^2 \gamma_0^2},~~~\gamma_0= \frac{66 N^{(0)}}{7 Q r_+^3}.
\eeq
So in the absence of a 3-cycle in M$_7$ flux quantisation is again rather easily achieved. If $\Sigma_3 \subset \text{M}_7$ does exist we find that
\beq
N^{(4)}= \sqrt{\frac{2 N^{(1)}}{77 \text{Vol}(\text{M}_7)}}\int_{\Sigma_3}\Phi_3,
\eeq
with the square root indicating that a tuning of $N^{(1)}$ will be necessary. For the example of S$^7/\mathbb{Z}_k$ we find
\beq
N^{(4)}_{\pm}=\pm \sqrt{\frac{24 N^{(1)}}{77 k}},
\eeq
which is again not automatically integer but permits tunings of $(N_1,k)$ that allow for this.

\subsubsection{Solution bounded between OF1 and OF1 with $F_0\neq 0$}
To realise OF1-planes at each of the endpoints of $[0,r_+]$ we tune
\begin{align}
\alpha_0&= \frac{r_+^3(\gamma+7\gamma-\Gamma)^2 \sigma_0}{120 \gamma},~~~\beta_0=\frac{r_+^2(\gamma+7\gamma^2-\Gamma)\sigma_0}{12},~~~\gamma_0=r_+\gamma \sigma_0,\nn\\[2mm]
\Gamma&\equiv\sqrt{\gamma(1+\gamma)(16+49 \gamma(1+\gamma))},
\end{align}
which gives a well defined solution for $\gamma \in \mathbb{R}/ [-1,0]$. We have that $N^{(8)}=\sigma_0$, so $\sigma_0$ must be integer. The charges of D0-branes and F1-strings are
\begin{align}
N^{(0)}&=\frac{Q r_+^4(1+2\gamma)\sigma_0}
{660\gamma(1+\gamma)}
\left(
-\Gamma\left(8+49\gamma(1+\gamma)\right)
+49\gamma(1+\gamma)
\left(3+7\gamma(1+\gamma)\right)
\right),\nn\\[2mm]
N^{(1)}&= Q\left(b_0\sigma_0+\frac{7}{720}r_+^4(\gamma+7\gamma^2-\Gamma)^2\sigma_0^2\right),
\end{align}
which we can solve in terms of $(Q,b_0)$ to achieve flux quantisation. When M$_7$ admits a 3-cycle to define a D4-brane charge, the latter takes the form
\beq
N^{(4)}= f(\gamma)
\sqrt{\frac{3N^{(0)}\sigma_0}{2\operatorname{Vol}(\text{M}_7)}}
\int_{\Sigma_3}\Phi_3,~~~f(\gamma)\equiv\sqrt{
\frac{\gamma(1+\gamma)(1+2\gamma)}
{
49\gamma(1+\gamma)\left(3+7\gamma(1+\gamma)\right)
-\Gamma\left(8+49\gamma(1+\gamma)\right)
}
}.
\eeq
While $f$ is a rather non-trivial function of $\gamma$, it is fairly easy to show that for $\gamma>0$ it ranges continuously between $f\sim 0.26$ and $f\to \infty$. As such, $N^{(4)}$ can be quantised for arbitrary values of the other charges. This is easiest to see by example: if we consider again S$^7/\mathbb{Z}_k$ we find that
\beq
f=\sqrt{2 N^{(0)}\sigma_0 k} \times \frac{q}{6}~~~\Rightarrow~~~N^{(4)}= q N^{(0)} \sigma_0
\eeq
which makes $N^{(4)}$  integer when $q\in \mathbb{Z}$. Note that since we must have $k\geq 2$ for a cycle to exist on which we can integrate this charge in the first place, we have $f\geq \frac{1}{3}$ which exceeds the lower bound of $f$. 

\subsection{Global solutions with source D8-branes}
Further solutions can be created by including D8-brane sources along the interior of $r$. This works in much the same manner as for Branch 1 in section \ref{sec:sourceD8s}. Such D8-branes occur when $h'''$ exhibits a discontinuity at a point where $(h,(h')^2,h'')$ are continuous, so that the metric and dilaton are continuous but  $dF_0$ has a delta function source.

If assume a solution starts at $r=0$ with either a D8/O8 or OF1 singularity, we can assume that the space is divided into cells of length 1 such that in the $\ell$'th cell one has $r\in [\ell,\ell+1]$ in which
\beq
h=\alpha_{\ell}+\beta_{\ell}(r-\ell)+\frac{1}{2}\gamma_{\ell}(r-\ell)^2+\frac{1}{6}\sigma_{\ell}(r-\ell)^3.
\eeq
As it is still that $(h,(h')^2,h'')$ must be continuous as one crosses between adjacent cells, the continuity conditions are identical to \eqref{eq:continuityconds}, and at $r=\ell$ one has a source of the form
\beq
dF_0= -\frac{1}{2\pi}(\sigma_{\ell-1}-\sigma_{\ell})\delta(r-\ell)dr.
\eeq
Taking \eqref{eq:caseIImet} and \eqref{eq:C72} to define $(B,C_7)$ in the $\ell$'th cell we find that the Bianchi identities are consistent with only D8-brane sources being present when the worldvolume flux of the D8-brane is assumed to be zero.   

Let us now construct two solutions with interior D8-branes. As ensuring metric positivity is considerably more complicated in this case, we focus on solution with a single D8-brane, bounded between singularities of the same type. As such, we will have only two cells, one with $\ell=0$ and one with $\ell=1$. In our first example we will solve the continuity constraints of\eqref{eq:continuityconds} with the $-$ sign and in the second with the $+$.

\subsubsection{Solution bounded between D8/O8 and D8/O8}
To have a D8/O8 singularity at $r=0$ we must tune the parameters in the 0'th cell as
\beq
\beta_0=\gamma_0=0,
\eeq
where $(\alpha_0,\sigma_0)$ are for now free parameters. We solve the continuity constraints at $r=1$ and ensure that there is a second D8/O8 at $r=2$ by fixing the parameters in the 1st cell as
\beq
\alpha_1=\alpha_0+\frac{1}{6}\sigma_0,~~~\beta_1=-\frac{1}{2}\sigma_0,~~~\gamma_1= \sigma_0,~~~\sigma_1=-\sigma_0.
\eeq
This means that we have a D8-brane source of charge $2\sigma_0$ at $r=1$. Metric positivity is ensured if $\Delta>0$ which in this case is ensured across $0<r<2$ provided that we tune
\beq
\alpha_0>\frac{2}{15}\sigma_0,~~~~ \alpha_0,\sigma_0>0.  \label{eq:bounds1}
\eeq
Moving on to flux quantisation, we have that $(N^{(8)}_{0},N^{(8)}_{1})$ are appropriately quantised provided that $\sigma_0=N^{(8)}\in\mathbb{N}$. The D0-brane charge requires an integral over the entire internal and is independent of $b_{0}$ $b_{1}$; we find that this can be set to any value by tuning
\beq
Q=\frac{66 N^{(0)}}{192 \alpha_0-\sigma_0}.
\eeq
We should have that $C_7$ shifts by a large gauge transformation as the D8-brane at $r=1$ is traversed, \textit{i.e.} we should have
\beq
\frac{1}{(2\pi)^7}\int \Delta C_7 = c^{(7)}_{1}(r=1)-c^{(7)}_{0}(r=1) = n\in\mathbb{Z},\label{eq:largeg}
\eeq
which requires that we tune
\beq
Q\left(b_0-b_1-\frac{10}{11}\alpha_0+ \frac{13}{132}\sigma_0\right)= n.
\eeq 
Finally, we can fix the charge of F1-strings between $r=0$ and $r=1$, $N^{(1)}_0$, to any value by setting
\beq
b_0=\frac{N^{(1)}}{Q\sigma_0},
\eeq
yielding  $N^{(1)}_0$ integer if $N^{(1)}$ is integer. We thus find that the charge of D0- branes $N^{(0)}$ and charges of  D8-branes enter those of  F1-strings as
\begin{align}
N^{(1)}_0= N^{(1)},~~~N^{(8)}_0=N^{(8)},~~~ N^{(1)}_1=(n+N^{(0)}) N^{(8)}-N^{(1)},~~~N^{(8)}_1=-N^{(8)},
\end{align}
When M$_7$ contains a 3-cycle on which we can define a D4-brane charge we find that
\beq
N^{(4)}= \sqrt{\frac{3N^{(0)}}{5(192\alpha_0-N^{(8)})\text{Vol}(\text{M}_7)}}N^{(8)} \int_{\Sigma_3}\Phi_3.
\eeq
However, we are totally at liberty to tune $\alpha_0$ such that this too is integer, provided that we are consistent with the bound of \eqref{eq:bounds1}. For the case of S$^7/\mathbb{Z}_k$ we find that we can quantise
\beq
 N^{(4)}_{\pm}= \pm N^{(4)},
\eeq
and obey this bound provided that $12 N^{(0)}N^{(8)}> 41 k (N^{(4)})^2$.

\subsubsection{Solution bounded between OF1 with $F_0=0$ and D8/O8}
In order to realise a solution that is bounded between OF1-planes with $F_0=0$ at $r=0$, a D8/O8 system at $r=r_+>1$ with a stack of $N^{(8)}$ D8-branes at $r=1$ we can tune
\begin{align}
\alpha_0&=\frac{3}{10}(r_+-1)(r_++1)^2N^{(8)},~~~\beta_0= -\frac{1}{2}(r_+^2-1)N^{(8)},~~~\gamma_0=(r_+-1)N^{(8)},~~~\sigma_0=0,\\[2mm]
\alpha_1&=\frac{1}{10}(r_+-1)(3r_+^2+r_++3)N^{(8)},~~~\beta_1=-\frac{1}{2}(r_+-1)^2N^{(8)},~~~\gamma_1=(r_+-1)N^{(8)},~~~\sigma_1=-N^{(8)},\nn
\end{align}
where $N^{(8)}>0$. The metric is then guaranteed to be positive in the interval $r\in (0,r_+)$. For $r\in [0,1)$ we find that the charges of D8-branes and F1-strings are given by 
\beq
N^{(8)}_0=0,~~~N^{(1)}_0=\frac{7}{20}Q(r_+^2-1)^2(N^{(8)})^2,
\eeq
the second of which we take as a definition of $Q$ to quantise $N^{(1)}_0$.
For $r\in(1,r_+]$ we find
\beq
N^{(8)}_1=-N^{(8)},~~~N^{(1)}_1= -\frac{1}{20}Q N^{(8)}(20b_1-7(r_+^2-1)^2 N^{(8)}),
\eeq
so we can also quantise $N^{(1)}_1$ by tuning $b_1$. In order for $C_7$ to shift by a large gague transformation as the D8-brane at $r=1$ is traversed, one should impose \eqref{eq:largeg}, which is equivalent to  
\beq
Q(b_0-b_1)=n+\frac{1}{30}Q(18 r_+^3+3 r_+^2-8 r_+-13)N^{(8)},
\eeq
which fixes $b_1$. The charge of D0-branes now becomes  
\beq
N^{(0)}= \frac{N^{(1)}_0 }{N^{(8)}} g(r_+),~~~g(r_+) \equiv \frac{123 r_+^3+ 303 r_+^2-93 r_+-53}{231(r_+-1)(r_++1)^2}.
\eeq
It is possible to show that $g\to \infty$ as $r_+\to 1$ and decreases to a minimum of $~ 0.53$ as $r_+\to \infty$. As such, we take $g= N^{(8)}p$ for $p\in \mathbb{N}$ which quantises $N^{(0)}$.  This time, assuming there is a 3-cycle on which $F_4$ has support, we are lead to
\beq
N^{(4)}= \sqrt{\frac{N^{(1)}_0}{ 154\text{Vol}(\text{M}_7)}}\int_{\Sigma_3}\Phi_3.
\eeq
It is not possible to quantise the D4-brane charge for generic values of $N^{(1)}_0$, though it can still be achieved for specific values. Of course this only matters if M$_7$ contains a 3-cycle in the first place.

\section*{Acknowledgements}
The work of NTM is funded by the Ram\'on y Cajal fellowship RYC2021-033794-I, he also acknowledges support of the Spanish  government grants MCIU-22-PID2021-123021NB-I00, MCIU-25-PID2024-161500NB-I00.

\appendix

\section{The roots of the quartic polynomial $\Delta$}
\label{roots}

Let us write
\begin{equation}
h = b_3 r^3 + b_2 r^2 + b_1 r + b_0,
\end{equation}
and
\begin{equation}
\Delta = a_4 r^4 + a_3 r^3 + a_2 r^2 + a_1 r + a_0,
\end{equation}
where 
\begin{align}
a_4 &= - \frac{48 b_3^2}{5}, \quad
a_3 = - \frac{64 b_2 b_3}{5}, \quad
a_2 = - \frac{28 b_2^2+12 b_1 b_3}{5}, \nn \\
a_1 &= - \frac{28 b_1 b_2}{5} + 12 b_0 b_3, \quad
a_0 = - \frac{12 b_1^2}{5} + 4 b_0 b_2.
\end{align}
We study the roots of $\Delta$, following \cite{Rees:1922}, by forming its reduced form (we consider the case $b_3 \neq 0$)
\begin{equation}
\Delta_{\text{reduced}} = x^4 + q x^2 + r x + s,
\end{equation}
where
\begin{equation}
q = \frac{3 b_1 b_3 - b_2^2}{12 b_3^2}, \quad
r = -\frac{5 (2 b_2^3 - 9 b_1 b_2 b_3 + 27 b_0 b_3^2)}{108 b_3^3}, \quad
s =  \frac{(3 b_1 b_3 - b_2^2)^2}{36 b_3^4},
\end{equation}
and $x = r + \frac{a_3}{4 a_4}$. The fact that 
\begin{equation}
s \geq \frac{q^2}{4}
\end{equation}
means that in order for $\Delta$ to have two distinct real roots between which it stays positive, as required by positivity of the metric, the discriminant of $\Delta$, Disc($\Delta$) has to be negative \cite{Rees:1922}.
\begin{align}
\text{Disc}(\Delta) = 
\frac{65536}{625} b_3^2 \left(343 b_2^6 - 3087 b_1 b_2^4 b_3 + 
   54 b_2^2 (159 b_1^2 + 50 b_0 b_2) b_3^2 \right.  \nn \\ 
  \left. - 243 b_1 (27 b_1^2 + 50 b_0 b_2) b_3^3 + 18225 b_0^2 b_3^4\right)\text{Disc}(h)
\end{align}
and
\begin{equation}
\text{Disc}(h) = b_1^2 b_2^2 - 4 b_1^3 b_3 + 18 b_0 b_1 b_2 b_3 - b_0 (4 b_2^3 + 27 b_0 b_3^2).
\end{equation}
We note that $s = q^2/4$ is realized when $3 b_1 b_3 - b_2^2 = 0$ and because Disc($h'$) $= 4 (b_2^2 - 3 b_1 b_3)$, this is the case where $h'$ has a double root.

\section{Uplift of the massless solution of Branch 2 to $d=11$}\label{eq:appendix2}
\label{uplift}

Uplifting the massless solution to eleven dimensions, yields the metric
\begin{align}
ds^2_{11} &= \frac{h}{\Delta^{\frac{2}{3}}}\left[\frac{e^{\frac{4}{3}\Phi_0}}{h''}(d\phi+C_1)^2 + \frac{\sqrt{12}e^{-\frac{2}{3}\Phi_0}h''}{8\sqrt{5}}ds^2(\text{AdS}_2)\right] \nn \\ 
&+ e^{-\frac{2}{3}\Phi_0} \Delta^{\frac{1}{3}}\left[\frac{25}{36m^2}\sqrt{\frac{12}{5}} ds^2({\rm M}_7)+ \frac{1}{8}\sqrt{\frac{5}{12}}\frac{h''}{h}dr^2\right],
\end{align}
where
\begin{equation}
e^{-\Phi_0} \equiv \frac{18}{25}\left(\frac{3}{5}\right)^{\frac{1}{4}}\frac{1}{c_0 L^3}, ~~~
dC_1 = -\sqrt{\frac{3}{5}}\frac{9h''}{25c_0 L^2}\text{vol}(\text{AdS}_2)
\end{equation}

$h$ is a quadratic function, $h = b_2 r^2 + b_1 r + b_0$, and so is $\Delta$. Positivity of the metric requires $\Delta \geq 0$, and since the coefficient of the quadratic term of $\Delta$ is negative, $-\frac{28}{5} b_2^2$, this necessitates $r$ to be restricted between the roots of $\Delta$. For $\Delta$ to have real roots, its discriminant has to be positive:
\begin{equation}
{\rm Disc}(\Delta) = - \frac{112}{5} b^2_2 \, {\rm Disc}(h) > 0.
\end{equation}
This in turn requires the discriminant of $h$, ${\rm Disc}(h)$, to be negative, that is, $h$ has no real roots.

Near a root of $\Delta$, the type of sigularity corresponds to that of an OM2-plane extended on AdS$_3$ formed as a S$^1$ fibration over AdS$_2$.

\end{document}